\documentclass[notitlepage]{article}

\usepackage{geometry}
\usepackage{fancyhdr}

\usepackage{amsmath}
\usepackage{mathtools}
\usepackage{hyperref}
\usepackage{authblk}
\usepackage{graphicx}
\usepackage{caption}
\usepackage{cite}

\usepackage{xcolor}
\definecolor{cyan}{HTML}{009bfa}
\definecolor{orng}{HTML}{e36f47}

\usepackage[square,numbers]{natbib}
\date{}

\begin{document}

\title{On the Effect of Nucleation Undercooling on Phase Transformation Kinetics}

\author[1,2]{Jos\'{e} Mancias \thanks{jose.mancias@tamu.edu}}
\author[1]{{Vahid} {Attari} \thanks{attari.v@tamu.edu} }
\author[1,3,4]{{Raymundo} {Arr\'{o}yave} \thanks{raymundo.arroyave@tamu.edu} }
\author[2]{{Damien} {Tourret} \thanks{damien.tourret@imdea.org} }

\affil[1]{Department of Materials Science and Engineering, Texas A\&M University, College Station, 77843, TX, USA}
\affil[2]{{IMDEA Materiales}, {{Calle Eric Kandel 2}, {Getafe}, {28906}, {Madrid}, {Spain}}}
\affil[3]{{J. Mike Walker '66 Department of Materials Science and Engineering}, {Texas A\&M University}, {{College Station}, {77843}, {TX}, {USA}}}
\affil[4]{{Wm Michael Barnes '64 Department of Industrial and Systems Engineering}, {Texas A\&M University}, {{College Station}, {77843}, {TX}, {USA}}}

\maketitle

\abstract{
We carry out an extensive comparison between Johnson-Mehl-Avrami-Kolmogorov (JMAK) theory of first-order phase transformation kinetics and phase-field (PF) results of a benchmark problem on nucleation. 
To address the stochasticity of the problem, several hundreds of simulations are performed to establish a comprehensive, statistically-significant analysis of the coincidences and discrepancies between PF and JMAK transformation kinetics. 
We find that PF predictions are in excellent agreement with both classical nucleation theory and JMAK theory, as long as the original assumptions of the latter are appropriately reproduced --- in particular, the constant nucleation and growth rates in an infinite domain.
When deviating from these assumptions, PF results are at odds with JMAK theory. 
In particular, we observe that the size of the initial particle radius $r_0$ relative to the critical nucleation radius $r^*$ has a significant effect on the rate of transformation.
While PF and JMAK agree when $r_0$ is sufficiently higher than $r^*$, the duration of initial transient growth stage of a particle, before it reaches a steady growth velocity, increases as $r_0/r^*\to 1$. 
This incubation time has a significant effect on the overall kinetics, e.g., on the Avrami exponent of the multi-particle simulations. 
In contrast, for the considered conditions and parameters, the effect of interface curvature upon transformation kinetics --- in particular negative curvature regions appearing during particle impingement, present in PF but absent in JMAK theory --- appears to be minor compared to that of $r_0/r^*$.
We argue that rigorous benchmarking of phase-field models of stochastic processes (e.g., nucleation) need sufficient statistical data in order to make rigorous comparisons against ground truth theories. In these problems, analysis of probability distributions is clearly  preferable to a deterministic approach.

}

%\keywords{Phase Transitions, Phase-Field Modeling, Johnson-Mehl-Avrami-Kolmogorov Theory, Nucleation, Growth}

\thispagestyle{fancy} 

%%%%%%%%%%%%%%%%%%%%%%%%%%%%%%%%%%%%%%%%%%%%%%%
\section{Introduction}\label{sec:intro}
%%%%%%%%%%%%%%%%%%%%%%%%%%%%%%%%%%%%%%%%%%%%%%%

First-order phase transformations are ubiquitous in materials science. They occur through the processes of nucleation and growth, and are often modelled using Johnson-Mehl-Avrami-Kolmogorov (JMAK) theory~\cite{kolmogorov_andrei_nikolaevitch_study_1937, avrami_kinetics_1939, avrami_kinetics_1940, avrami_granulation_1941, william_a_johnson_reaction_1939}.
This simple, yet rigorous, mean-field approach describes, for an infinite domain undergoing an isothermal transformation with no composition change, the evolution of the transformed fraction, $Y$, as a function of time, $t$, in a system with $d$ dimensions as \cite{christian_theory_2002, cahn_theory_1962}
\begin{align} \label{avrami_1}
    Y(t) = 1 - \exp\left(- \bar{Y}(t)\right) , 
\end{align}
where 
\begin{align} \label{avrami_intg} 
    \bar{Y}(t) = K_d \int^t_{0} J(\xi) \left(\int^t_\xi V(\zeta) d\zeta \right)^d  d\xi
\end{align}
represents the extended volume of potentially overlapping particles. 
The time-dependent nucleation rate, $J$, and growth rate, $V$, i.e., the normal velocity of the particle-matrix interface, are both integrated over time.
$K_d$ is a geometric factor that represents the volume of a $d$-dimensional particle of unit radius, i.e., $K_1=2$, $K_2=\pi$, and $K_3=4\pi/3$ in one, two, and three dimensions, respectively. On the one hand, for constant rates of nucleation, $\bar J$, and growth, $\bar V$, the integration of equation~\eqref{avrami_intg} gives:
\begin{align} \label{eq:avrami_cont_nucl}
    \bar Y(t) = \frac{K_d}{d+1}\,\bar J \,\bar V^d \,t^{d+1} .
\end{align}
On the other hand, if particle nucleation occurs in one short initial burst, e.g., due to the presence of preferential nucleation sites such as grain boundaries, one may consider a case commonly called ``site saturation'' with $J(0)=N$, with $N$ the number of nuclei and $J(t\neq0)=0$, thus leading to: 
\begin{align} \label{eq:site_saturation}
    \bar Y(t) = K_d N\,\bar J \,\bar V^d \,t^d .
\end{align}
Hence, rates of transformations are commonly described in a general form as:
\begin{align} 
    Y(t) = 1 - \text{exp}(-kt^n) ,
\end{align}
where $n$ is referred to as the Avrami exponent. 
This exponent can be conveniently evaluated from the slope of a plot of $\log[-\log(1-Y)]$ versus $\log(t)$. 
Historically, transformations have been analyzed and often classified based on their Avrami exponent, even for cases that span beyond its rigorous scope of applicability \cite{christian_theory_2002}.
The equation has been used across a broad range of disciplines, e.g., in epidemiology to model the spread of disease in a population \cite{avramov_kinetics_2007}. 

The simplicity of the JMAK theory comes at the expense of some strong assumptions. 
Importantly, the approach relies on the key principle of statistical homogeneity, in time and space, of nucleation events – an assumption that is approached in many cases of first order phase transformations. 
The JMAK equation is the statistical solution for random nucleation events with constant rates for nucleation and growth rate in an infinite domain. 
It is also valid for a common time-dependent growth rate or anisotropic convex particles with a parallel orientation~\cite{granasy_phase-field_2014}. 

Being an exact statistical solution, the JMAK equation can be obtained following different pathways. 
Johnson and Mehl \cite{william_a_johnson_reaction_1939} and Avrami \cite{avrami_kinetics_1939, avrami_kinetics_1940, avrami_granulation_1941} computed a transformed fraction ignoring impingement, then corrected to discard regions counted multiple times.
Kolmogorov \cite{kolmogorov_andrei_nikolaevitch_study_1937} performed a time integration of the probability of transformation of an untransformed point.
The original theory considers a uniform nucleation rate across the entire domain, even in already transformed regions, thereby introducing the constructs of ``phantom nuclei’’ and ``extended volume transformed’’, and then applies an exact correction to adjust for overlapping nucleation and growth within the already transformed region. 

This stochastic correction for the multiple integrations over the extended volume fails for finite domain sizes, which results in spatial stochastic inhomogeneity.
In such cases, the more general ``time cone’’ approach \cite{jackson_dynamics_1974, cahn_time_1996, balluffi_kinetics_2005} allows modeling time-dependent rates of nucleation and growth in finite, inhomogeneous, and evolving domains. 
This theory, proposed by Cahn \cite{cahn_time_1996} extending an original idea by Jackson \cite{jackson_dynamics_1974}, considers a time-augmented space of dimension $d+1$. 
Therein, a cone apex represents a spatiotemporal location, and the cone itself represents the prior spatiotemporal nucleation locations that would have lead the apex location to undergo the transformation.
The integration of the Poisson-distributed -- because of stochastic independence -- probability of prior nucleation events within the cone leads to a general formulation, which reduces to the original JMAK theory under the assumptions of homogeneous nucleation and constant nucleation and growth rates.

With advances in computational models -- in particular using phase-field (PF) approaches \cite{chen_phase-field_2002, boettinger_phase-field_2002, moelans_introduction_2008, tourret_phase-field_2022, granasy_phase-field_2014} -- and high-performance hardware allowing simulations of statistically relevant sizes, first-order transformation kinetics have been simulated and directly compared to JMAK theory \cite{jou_comparison_1997, simmons_microstructural_2004, granasy_phase-field_2014}. These studies consistently exhibited a good agreement with the exact theory within the bounds of its original assumptions, as well as some expected deviations when breaching them. 
Such deviations are, for instance, related to (i) the particles initial slower growth rates  \cite{jou_comparison_1997}, (ii) exceedingly low or high nucleation rates \cite{jou_comparison_1997}, or (iii) diffusion-mediated growth leading to a so-called ``soft impingement’’ when diffusion plays a key role \cite{ simmons_microstructural_2004}.
One key interest of the numerical simulations is the capacity to explore the extent of the breakdown of the JMAK-like kinetics as assumptions are relaxed.

In this article, we take another look at transformation kinetics involving concurrent nucleation and growth, using PF simulations.
To do so, we use a computationally efficient implementation of a benchmark PF simulation (see section~\ref{sec:methods}), and perform a statistical analysis over hundreds of simulations, recognizing the stochastic nature of the problem. 
Moreover, we discuss the effect of the driving force, e.g., undercooling, in relation to the initial particle size upon the potential deviation from theoretical values of the Avrami exponent.
Section~\ref{sec:methods} discusses the methods, including the nucleation benchmark problem, the implementation, and further post-processing. Phase-field results and relevant comparisons with JMAK theory are presented and discussed in sections~\ref{sec:results} and~\ref{sec:discussion}, respectively. We conclude in section \ref{sec:conclusion}, arguing that only an explicit consideration of the statistical distribution of simulation observables provides a full understanding of the transformation kinetics. 

%%%%%%%%%%%%%%%%%%%%%%%%%%%%%%%%%%%%%%%%%%%%%%%
\section{Methods}\label{sec:methods}
%%%%%%%%%%%%%%%%%%%%%%%%%%%%%%%%%%%%%%%%%%%%%%%

%%%%%%%%%%%%%%%%%%%%%%%%%%%%%%%%%%%%%%%%%%%%%%%
\subsection{Phase-Field Benchmark Problems} \label{sec:pf-benchmark}

The Center for Hierarchical Materials Design (CHiMaD) and the National Institute of Standards and Technology (NIST) have developed a series of PF benchmark problems, hosted on the PFHub website \cite{wheeler_pfhub_2019}.
Therein, different benchmark problems are described, based on simple prototypical models, where users can post and compare their solutions. 
These benchmarks allow PF modelers to test their codes and for novices to learn from simple problems \cite{jokisaari_benchmark_2017}. 

One recently published benchmark problem, considered in this article, focuses on nucleation \cite{wu_phase_2021}.
The problem considers an undercooled liquid undergoing a first-order phase transformation through nucleation and growth mechanisms.
Classical nucleation theory (CNT) and JMAK theory are used to verify the simulation results in terms of critical nucleation radius and transformation rates for both continuous nucleation and site-saturation cases. 
The key features of the model and benchmark simulation cases are described in the following subsections, as well as the numerical solving and post-processing techniques specific to the current article.

%%%%%%%%%%%%%%%%%%%%%%%%%%%%%%%%%%%%%%%%%%%%%%%
\subsection{Phase-Field Model} \label{sec:pf-model}

The considered PF model involves a single non-conserved field order parameter ($0\leq\phi\leq 1$) describing the states of matter under isothermal undercooled conditions.
The total free energy of the heterogeneous system is described as:
\begin{equation} \label{eq:free_energy}
    \mathcal{F}(\phi) = \int_V \left[ \frac{\epsilon^2}{2}(\nabla\phi)^2 + wg(\phi) - \Delta f p(\phi) \right] dV
\end{equation}
\noindent
where $\epsilon$ is the gradient energy coefficient, $w$ specifies the barrier height of phase transition, and $\Delta f$ is the driving force for transformation (e.g., undercooling). $g(\phi)=\phi^2(1-\phi)^2$ is a double-well function with two minima at $\phi = 0$ (phase/state 1) and $\phi = 1$ (phase/state 2), and $p(\phi)=\phi^3(10-15\phi+6\phi^2)$ is an interpolation function satisfying $p(0) = p'(0) = p'(1) = 0$ and $p(1) = 1$. 
The Allen-Cahn equation \cite{AllenCahn1979} describes the time evolution of $\phi$ as:
\begin{equation} \label{eq:allen_cahn}
    \frac{\partial \phi}{\partial t} = -M \frac{\delta \mathcal{F}}{\delta \phi} = M [\epsilon^2 \nabla^2 \phi - wg'(\phi) + \Delta f p' (\phi)],
\end{equation}
where $M>0$ is the PF mobility. 
The interface width and excess free energy are thus $l =\sqrt{\epsilon^2/w}$ and $\gamma = \sqrt{\epsilon^2w}/(3\sqrt{2})$, respectively. 

The two-dimensional (2D) domain is initialized with $\phi=0$, and nuclei of radius $r_0$ are then added as circular regions by setting 
\begin{equation}\label{initialCondition}
\phi(r_i) = \frac{1}{2} \left[1-\tanh\left(\frac{r_i-r_0}{l\sqrt{2}}\right) \right], 
\end{equation} 
where $r_i=\sqrt{(x-x_i)^2+(y-y_i)^2}$ is the radial distance to the center $(x_i,y_i)$ of nucleus $i$.
When seeded nuclei overlap with a transformed region (e.g., in the case of JMAK ``ghost nucleation'' events), their individual contributions to the field $\phi$, per Eq.~\eqref{initialCondition}, are added while keeping the field bounded to $\phi\leq 1$.

Following CNT, the excess free energy of a circular nuclei of radius $r$ is
\begin{equation}
    \Delta G (r) = 2\pi r \gamma - \pi r^2 \Delta f
\end{equation}
and the critical nucleation radius, given by ${\partial \Delta G(r)}/{\partial r}=0$ is 
\begin{equation}
    r^* = \gamma/\Delta f ,
\end{equation}
for a corresponding free energy $\Delta G^* = \pi \gamma^2/\Delta f$.
Any particle with $r_0<r^*$ is thus expected to shrink, while any particle with $r_0>r^*$ is expect to grow in order to reduce the overall free energy of the system.

By normalizing space with respect to the interface width $l$ and time with respect to a characteristic time $\tau = 1/(Mw)$, the dimensionless form of the equations \eqref{eq:free_energy} and \eqref{eq:allen_cahn} reads
\begin{equation}\label{FreeEnergyEq}
    \widetilde{\mathcal{F}} (\phi) = \int \left[ \frac{1}{2} (\widetilde{\nabla} \phi)^2 + g(\phi) - \widetilde{\Delta f} p(\phi) \right] d \widetilde{V}
\end{equation}
\begin{equation}\label{eq:adim_allen_cahn}
    \frac{\partial\phi}{\partial\Tilde{t}} = \widetilde{\nabla}^2\phi - g'(\phi) + \widetilde{\Delta f}p'(\phi)
\end{equation}
where $\widetilde{\Delta f} = \Delta f / w$. 
In this form, new nuclei are seeded with 
\begin{equation}\label{eq:adim_seed}
\phi(\tilde r_i) = \frac{1}{2} \left[1-\tanh\left(\frac{\tilde r_i-\tilde r_0}{\sqrt{2}}\right) \right].
\end{equation} 
and the critical nucleation radius is 
\begin{equation}\label{eq:adim_cnt}
    \tilde r^* = \frac{1}{3\sqrt{2}} \frac{1}{\widetilde{\Delta f}}.
\end{equation}
The non-dimensional form of Eqs~\eqref{eq:adim_allen_cahn}-\eqref{eq:adim_cnt} are numerically implemented. 
In the remainder of this article, for clarity (and in order to be consistent with the original paper \cite{wu_phase_2021}), we refer exclusively this dimensionless form of the equations, but drop the tilde notation on all symbols, e.g., writing simply $\Delta f$, $r_0$, and $r^*$ in place of $\widetilde{\Delta f}$, $\tilde r_0$, and $\tilde r^*$.

%%%%%%%%%%%%%%%%%%%%%%%%%%%%%%%%%%%%%%%%%%%%%%%
\subsection{Nucleation Benchmark Simulations} \label{sec:nucleation_simulations}

Based on the PF model described above, the benchmark study is composed of three parts. 
\emph{Problem~I} involves a single seed with a radius $r_0$ close to the critical radius of nucleation $r^*$.
The results are compared to classical nucleation theory (Eq.~\eqref{eq:adim_cnt}).
\emph{Problem~II} and \emph{Problem~III} represent site-saturation and continuous nucleation, respectively, of several particles.
The results are directly compared to JMAK theory, i.e., respectively Eqs~\eqref{eq:site_saturation} and \eqref{avrami_intg}.
Table~\ref{simulationParams} summarizes parameter values (in dimensionless units of $l$ and $\tau$) of domain size, driving force ($\Delta f$), corresponding critical radius ($r^*$), initial nuclei size ($r_0$), and total number of seeds ($N$) for all cases.

The first problem (\emph{Problem~I}) involves three sub-problems where one side of various sizes (i.e., $r_0>r^*$, $r_0<r^*$, and $r_0=r^*$) is added at the center of the domain. This problem serves as a comparison to CNT, verifying that a seed with $r_0<r^*$ shrinks, and a seed with $r_0>r^*$ grows. 
The seed with $r_0=r^*$ is expected to stagnate in size until pushed towards either of these two outcomes. The fact that the radius does not remain at its initial value may be attributed to the diffuse interface nature of the PF approach, while CNT considers a sharp interface. Hence, the two are only expected to agree when the interface width is negligible. This effect is also possibly accompanied by potential numerical (e.g. discretization) effects affecting the outcome (growth or shrinkage).
In the interest of discussing the effect of the initial radius $r_0$ on the initial growth kinetics, in addition to the original values of $r_0/r^*=0.99$, 1.00, and 1.01 from the original benchmark definition, here we also show the results for $r_0/r^*=1.1$ and 2.0.
In order to facilitate the discussion, we refer to this ratio as $\rho=r_0/r^*$.

\emph{Problem~II} considers multiple seeds ($N=25$), all appearing at time $t = 0$, placed at random locations, and allowed to grow until the entire domain is fully transformed. 
\emph{Problem~III} is larger in size, with a higher number of nuclei ($N=100$), the key difference being that they are added at random times to replicate a constant nucleation rate. 
In both cases, new seeds are initialized with a unique radius $r_0=2.2$, whereas $r^*$ is set equal to either 1.0 or 2.0. While the original benchmark definition considers separate values of $\Delta f$, and hence of $r^*$ for \emph{Problem~II} and \emph{Problem~III}, here we perform both cases for both values of $\Delta f$, namely with $\rho=1.1$ or 2.2 (see Table~\ref{simulationParams}).
In all cases, periodic boundary conditions need to be applied in order to emulate the JMAK assumptions, in particular that of infinite domain and spatiotemporal homogeneity of nucleation and growth, which are violated in a finite size system~\cite{cahn_time_1996}.

\begin{table}[!h]
\centering
\caption{Simulation parameters~\cite{wu_phase_2021}.}\label{simulationParams}%
\begin{tabular}{@{}lllllll@{}}
\hline
Problem \# & Domain Size & $\Delta f$ & $r^*$ & $\rho=r_0/r^*$ & N\\
\hline
Problem I    & 100$\times$100 & $\sqrt{2}/30$  & 5 & $0.99$, $1.00$, $1.01$, $1.1$, $2.0$ & 1  \\
Problem II    & 500$\times$500 & $\sqrt{2}/6$  & 1 & 2.2 & 25 \\
            & 500$\times$500 & $\sqrt{2}/12$ & 2 & 1.1 & 25 \\
Problem III   & 1000$\times$1000 &$\sqrt{2}/6$ & 1 & 2.2 & 100 \\
            & 1000$\times$1000 &$\sqrt{2}/12$  & 2 & 1.1 & 100  \\
\hline
\end{tabular}
\end{table}

%%%%%%%%%%%%%%%%%%%%%%%%%%%%%%%%%%%%%%%%%%%%%%%
\subsection{Numerical Implementation}\label{sec:implementation}

We solve the Allen-Cahn equation \eqref{eq:adim_allen_cahn} using centered finite differences and an explicit Euler time stepping scheme. 
Following a preliminary convergence analysis of our implementation, we selected a mesh size of $0.4$, used in all simulations. 
The model was implemented in Julia, a general purpose high-level, high-performance, dynamic programming language \cite{bezanson_julia_2012}.
We make use of the CUDA (Compute Unified Device Architecture) package (cuda.jl) \cite{besard2018juliagpu}, which enables code acceleration via multithreading on Nvidia graphics processing units (GPUs). 
While computational efficiency and scaling analysis fall beyond the scope of the current paper, we noticed a speed-up factor of 20 to 50 between Julia-GPU (NVIDIA RTX 2080) and Julia-serial (Intel Xeon Gold 6130) execution times of the three benchmark problems.

The computational efficiency afforded by the GPU-parallelized code constitutes a great asset in addressing stochasticity in the results of the benchmark study.
Indeed, nucleation events being randomly located in time (\emph{Problem~III}) and space (\emph{Problems~II \& III}), greatest statistical accuracy can be gained by running a large number of simulations. 
Hence, we preformed 300 PF simulations, each of them with a different random number seed, for each value of $\rho$ in both \emph{Problems~II} and \emph{III}. 
This study is also intended to provide a greater statistical insight into the deviation of observed behaviors (in particular transformation kinetics) for a high yet finite number of particles ($N=25$ or $100$) compared to JMAK theory.

%%%%%%%%%%%%%%%%%%%%%%%%%%%%%%%%%%%%%%%%%%%%%%%
\subsection{Post Processing}\label{sec:postproc}

In all parts of the benchmark, the transformed volume fraction and free energy, as well as its individual components, are tracked as a function of time.
In \emph{Problem~I}, the main outcome of the simulation is whether the particle grows or shrinks, and at which rate, such that a simple plot of the transformed volume fraction is sufficient to conclude and discuss the results. 

In \emph{Problems~II} and \emph{III}, the Avrami exponent $n$ can be extracted from a modified log-log plot of the transformed fraction $Y$ versus time $t$, namely plotting $\log[-\log(1-Y(t))]$ against $\log(t)$, which results in a straight line of slope $n$ for a perfect match to JMAK theory. 
Given the finite number of seeds, performing this linear fit for a given time range may lead to strong deviations from a linear behavior, and hence high error in the estimation of $n$, particularly in the case of late nucleation of the first seed in a simulation. 
Therefore, we elected to fit over a transformed fraction range rather than a time range, which we observed to exhibit more consistency.
Hence, we perform a least-square fit for the range $-2\leq\log[-\log(1-Y(t))]\leq0$, which correspond to a transformed fraction between $0.0228$ and $0.9$, hence representing a substantial portion of the transformation range.
Taking advantage of the hundreds (300) of simulations preformed for each configuration, results of time-dependent volume fractions and Avrami exponents can both be visualized (\ref{sec:results}) and discussed (\ref{sec:discussion}) in terms of their statistical distribution.

Finally, in order to assess the effect of interface curvature upon the transformation kinetics, we calculated and monitored it in time and space during PF simulations. 
This is of particular importance because of regions of negative curvature that develop during the impingement of particles. These regions are naturally captured in PF simulations, but absent from JMAK theory.
The local interface curvature can be directly obtained from the phase field $\phi$. 
Specifically, from the definition of the local curvature $\kappa = \nabla \cdot \mathbf{n}$, where $\mathbf{n}= \nabla \phi / |\nabla \phi|$ is the interface normal vector, the local interface curvature is directly calculated in the entire domain as \cite{vakili_numerical_2017}
\begin{equation}\label{eq:finalCurvature}
    \kappa = \frac{1}{|\nabla \phi|} \left( \nabla^2 \phi - \frac{\nabla \phi \cdot \nabla |\nabla \phi|}{|\nabla \phi|}\right) .
\end{equation}
For visualization purposes, when plotting the curvature maps in section~\ref{sec:results}, we filter out bulk regions (where $|\nabla\phi|=0$) by actually plotting the field $\kappa\exp\{-(\phi-1/2)^2/a\}$, which peaks at the value $\kappa$ at the interface location ($\phi=1/2$) and vanishes within a narrow distance from the interface, with a filtering parameter $a = 0.01$.

%%%%%%%%%%%%%%%%%%%%%%%%%%%%%%%%%%%%%%%%%%%%%%%
\section{Results}\label{sec:results}
%%%%%%%%%%%%%%%%%%%%%%%%%%%%%%%%%%%%%%%%%%%%%%%

%%%%%%%%%%%%%%%%%%%%%%%%%%%%%%%%%%%%%%%%%%%%%%%
\subsection{Problem I}\label{sec:resu:part1}

\begin{figure}[h!]%
\centering
\includegraphics[width=0.7\textwidth]{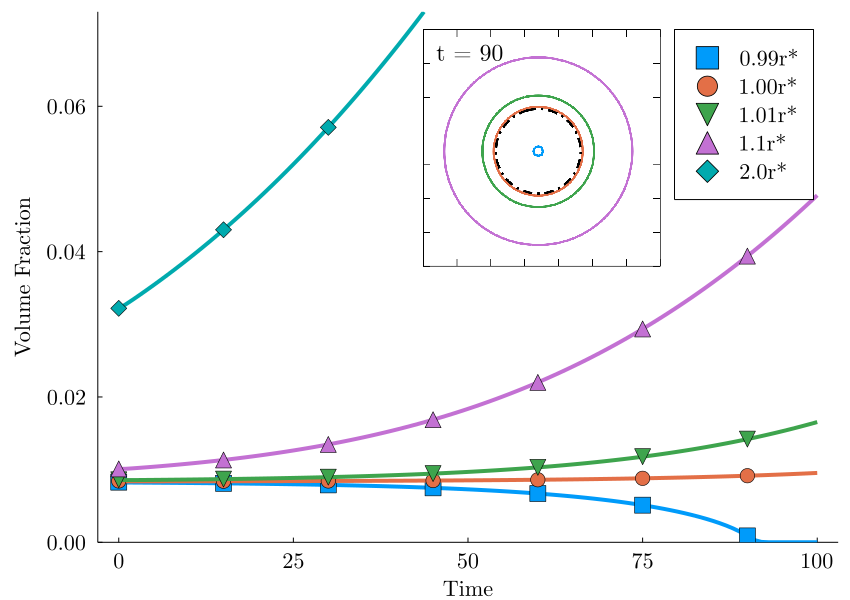}
\captionsetup{width=0.9\textwidth}
\caption{
Phase-field results for transformed fraction of a single seed (\emph{Problem~I}) versus time.
The inset shows the interface ($\phi=0.5$) at $t=90$, for the four cases with $\rho\leq1.1$, as solid lines and the initial radius as black dash-dotted line.
}\label{part1}
\end{figure}

Figure~\ref{part1} shows the results of simulations for \emph{Problem~I}.
The PF results show excellent agreement with CNT. 
The seed smaller than the critical radius shrinks, while those larger than the critical radius grow. 
At $r_0=r^*$, the radius should theoretically stagnate. 
However, the problem discretization and its numerical solution inevitably leads to small approximations (e.g., rounding errors) that ultimately accumulate and lead to the seed growing or shrinking. 
In the case of our implementation, Fig.~\ref{part1} shows that the particle grows with a sluggish initial kinetics.
In most cases, the growth or shrinkage of particles comes after a transient incubation time, which is longer as $r_0$ is closer to $r^*$.

%%%%%%%%%%%%%%%%%%%%%%%%%%%%%%%%%%%%%%%%%%%%%%%
\subsection{Problem II}\label{sec:resu:part2}

\begin{figure}[h!]%
\centering
\includegraphics[width=0.85\textwidth]{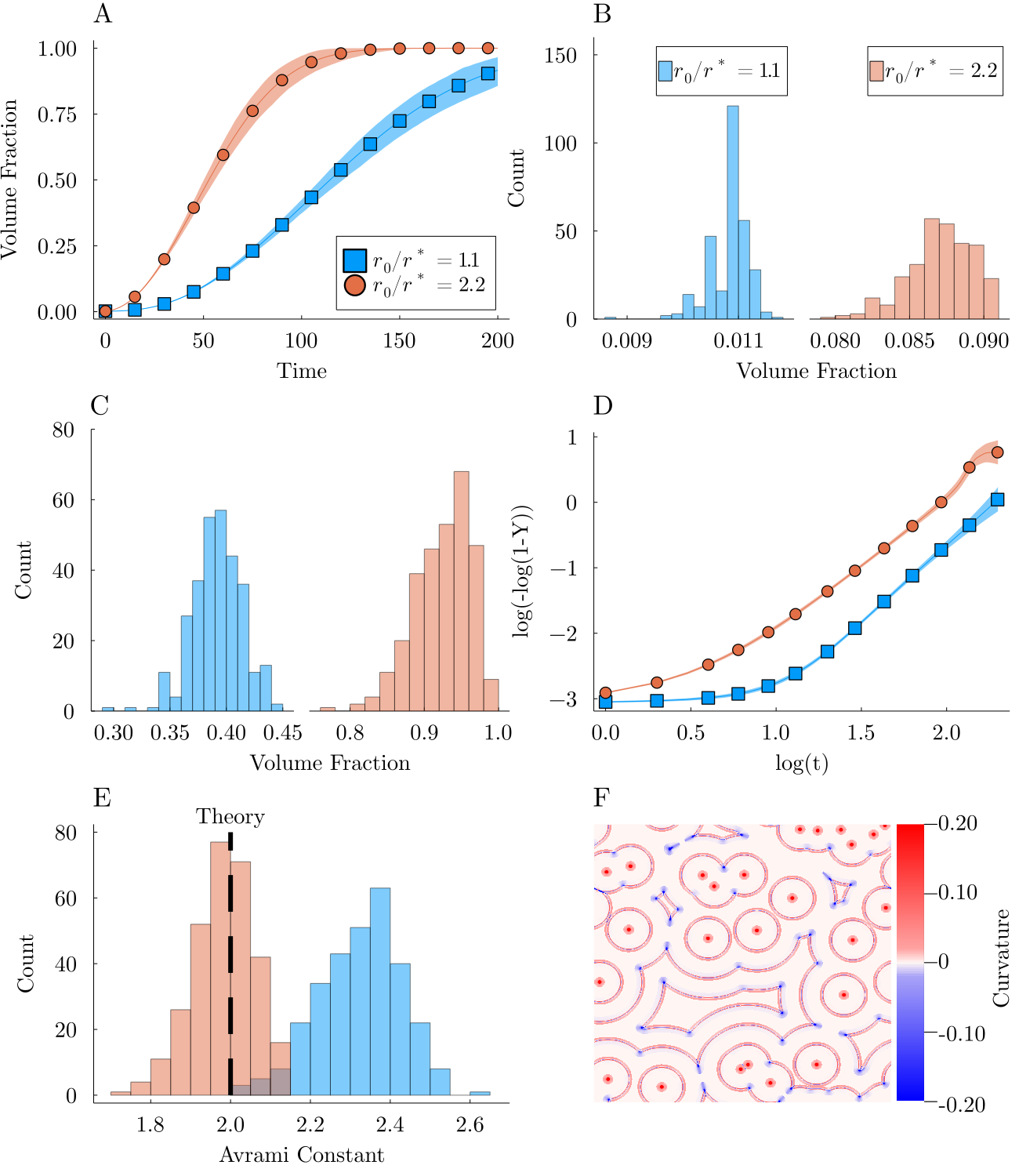}
\captionsetup{width=0.9\textwidth}
\caption{
Phase-field results of the site-saturation simulations (\emph{Problem~II}) with two distinct critical radius values $\rho = 1.1$ \raisebox{1mm}{\fcolorbox{black}{cyan}{\null}} and $\rho = 2.2$ \raisebox{1mm}{\fcolorbox{black}{orng}{\null}} combining 300 separate instances for each with different random location of the 25 initial seeds: 
(A) Transformed fraction versus time, showing global average (line and symbols) and statistical distribution (shaded background), 
(B) Volume fraction distribution at $t = 20$, 
(C) Volume fraction distribution at $t = 100$, 
(D) Avrami log-log plot, showing global average (line and symbols) and statistical distribution (shaded background), 
(E) Distribution of Avrami exponents fitted individually to the 300 simulations, 
(F) Spatiotemporal map showing the curvature of the particles at the interface as they grow at $t = 0, 40, 80, 120, 160, 200$ for one simulation with $\rho=1.1$.
}\label{part2}
\end{figure}

Figure~\ref{part2} shows the results for \emph{Problem~II} for two values of $\Delta f$, corresponding to $\rho=1.1$ and 2.2 (Table~\ref{simulationParams}). 
The 25 seeds of radius $r_0$ are placed randomly in the domain at the beginning of the PF simulation. 
As expected, since $r_0>r^*$, all particles grow, and eventually the entire domain is transformed (within the simulated time for $\rho=2.2$ and later for $\rho=1.1$). 
Due to the stochasticity in the random placement of the seeds at $t=0$, the 300 independent instances with different random locations lead to a statistical distribution of results, shown as shaded regions (A, D) or histograms (B, C, E) in Fig.~\ref{part2}.

Panel A of Fig.~\ref{part2} shows the time evolution of the transformed volume fraction. 
Panels B and C show the volume fraction distribution at times $t = 20$ and $t = 100$, respectively. 
Panel D shows the ``Avrami plot'' of $\log[-\log(1-Y(t))]$ versus $\log(t)$. 
The slope of its linear region corresponds to the Avrami exponent $n$, of which the corresponding distribution appears in panel E.
The least-square fit between $-2$ and $0$ on the y-axis (see section~\ref{sec:postproc}) leads to average slopes $n = 1.98$ and $2.32$, respectively for $\rho = 2.2$ and $1.1$, while a value of 2.0 is expected from JMAK theory (Eq.~\eqref{eq:site_saturation}).

Finally, panel F illustrates the system evolution for one typical simulation for $\rho=1.1$, in the form of a map of curvature $\kappa$ in the interface region (see section~\ref{sec:postproc}) for six snapshots from $t=0$ to $t=200$ with steps of 40.
This spatiotemporal map exhibits the 25 initial seeds (highest curvature localized at the initial grain centers), as well as region of negative curvature when the particles impinge on each other. 

%%%%%%%%%%%%%%%%%%%%%%%%%%%%%%%%%%%%%%%%%%%%%%%
\subsection{Problem III}\label{sec:resu:part3}

\begin{figure}[h!]%
\centering
\includegraphics[width=0.85\textwidth]{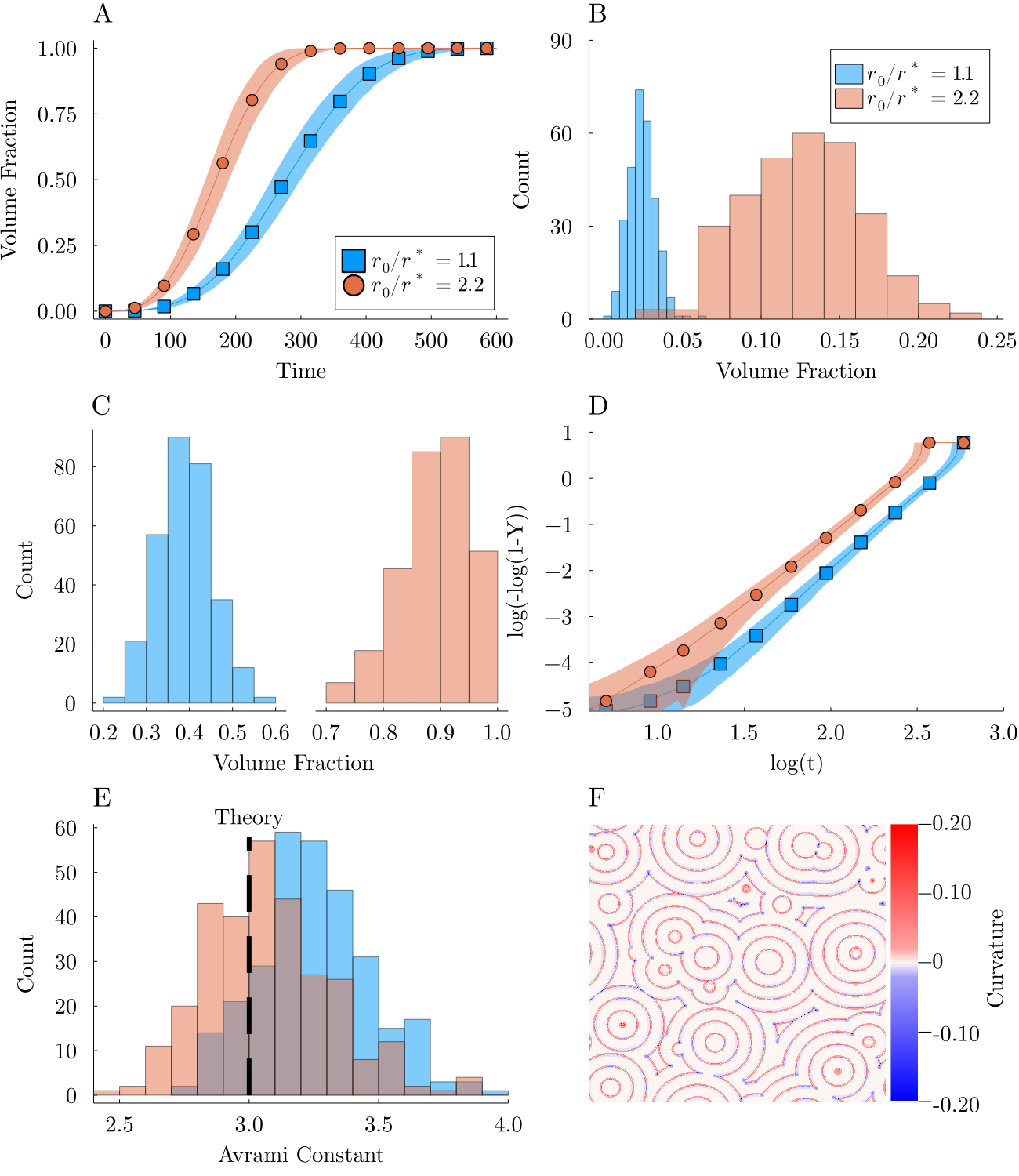}
\captionsetup{width=0.9\textwidth}
\caption{
Phase-field results of the continuous nucleation simulations (\emph{Problem~III}) with two distinct critical radius values $\rho = 1.1$ \raisebox{1mm}{\fcolorbox{black}{cyan}{\null}} and $\rho = 2.2$ \raisebox{1mm}{\fcolorbox{black}{orng}{\null}} combining 300 separate instances for each with different random location of the 25 initial seeds: 
(A) Transformed fraction versus time, showing global average (line and symbols) and statistical distribution (shaded background), 
(B) Volume fraction distribution at $t = 100$, 
(C) Volume fraction distribution at $t = 250$, 
(D) Avrami log-log plot, showing global average (line and symbols) and statistical distribution (shaded background), 
(E) Distribution of Avrami exponents fitted individually to the 300 simulations, 
(F) Spatiotemporal map showing the curvature of the particles at the interface as they grow at $t = 0, 50, 100, 150, 200, 250, 300$ for one simulation with $\rho=1.1$.
}\label{part3}
\end{figure}

Figure~\ref{part3} shows the results for \emph{Problem~III} for $\rho=1.1$ and 2.2. 
Figure panels are similar to that of Fig.~\ref{part2} described in section~\ref{sec:resu:part2}.
Once again, Fig.~\ref{part3} compiles the results of 300 independent instances of the problem with different randomly generated spatiotemporal locations of the 100 nucleation events in each simulation. 
The average Avrami exponents from PF are $n = 3.08$ and $n = 3.26$ respectively for $\rho = 2.2$ and $\rho = 1.1$, whereas the theoretical value from JMAK theory (Eq.~\eqref{eq:avrami_cont_nucl}) is $n = 3$. 

%%%%%%%%%%%%%%%%%%%%%%%%%%%%%%%%%%%%%%%%%%%%%%%
\section{Discussion}\label{sec:discussion}
%%%%%%%%%%%%%%%%%%%%%%%%%%%%%%%%%%%%%%%%%%%%%%%

Results of \emph{Problem~I} (Fig.~\ref{part1}) show the excellent agreement between PF simulations and CNT, in terms of critical nucleation radius for $r^*$ a given driving force $\Delta f$.
The simulation with $r_0=r^*$ slowly takes off and leads to the particle growth, which we attribute to the discretization and computational artifacts (e.g., rounding errors) accumulating until the particle radius gets either lower or higher than its initial value.
An interesting observation from Fig.~\ref{part1} is the relation between the incubation time $\tau_0$, i.e., the time it takes before the particle approaches a steady velocity.
PF results show that the transient growth period is shorter as the ratio $\rho=r_0/r^*>1$ increases.
This behavior is nothing surprising, as one would also expect that $\tau_0\to \infty$ as $r_0\to r^*$.
This initial transient has important consequences on the overall kinetics of the many-particle simulations of \emph{Problems~II} and \emph{III}.

The analysis of \emph{Problems~II} (Fig.~\ref{part2}) and \emph{III} (Fig.~\ref{part3}) clearly shows the dependence of the resulting growth kinetics (e.g., Avrami exponent) upon the combination of driving force (through the critical nucleation radius) and initial particle radius. 
Both cases exhibit a significantly higher growth rate (A), and a corresponding lower Avrami exponent (E), when the ratio of initial radius to critical radius is larger. Statistically, distributions of both transformed fraction at a given time (B,C) and that of Avrami exponents (E) are broader, which we attribute to the higher degree of stochasticity found in \emph{Problem~III}, specifically the random nucleation times.

The dependence of Avrami exponent ($n$) upon the ratio ($\rho$) of nuclei radius to critical radius is to a great extent a consequence of the change in incubation time observed in \emph{Problem~I} (Fig.~\ref{part1}). 
Large incubation times (as $\rho\to 1$) slow down the growth of individual particles, and consequently the average transformation kinetics. 
Another way to interpret these results is that, for a given initial particle radius $r_0$, a higher driving force (e.g., an higher undercooling), inversely proportional to $r^*$ (Eq.~\eqref{eq:adim_cnt}), reduces the incubation time.
Since JMAK theory assumes a constant growth rate, i.e., a negligible incubation time, a higher $\rho$ (i.e., a higher $\Delta f$) provides a better agreement between PF and JMAK kinetics.
These trends are a direct consequence of JMAK assumptions, in agreement with prior discussions of the theory based on analytical arguments \cite{cahn_time_1996} or simulation results \cite{jou_comparison_1997}.
One additional technical (post-processing) aspect worth mentioning is our observation that the extracted Avrami exponent is strongly sensitive on the choice of linear fitting range when using a time range -- hence our choice to use a transformed fraction range (see section~\ref{sec:postproc}), which seemed to lead to more consistent results. 

In addition to the effect of the incubation time, we also explored the effect of interface curvature on the overall transformation kinetics, in particular questioning the effect of negative curvature regions formed during the impingement of particles (see panel F of figures~\ref{part2} and~\ref{part3}). 
This effect, naturally integrated into PF simulations, is absent from the mean-field curvature-agnostic JMAK theory. 
However, while it seems most likely that these negative curvature regions affect the kinetics of transformation in the PF simulations, their effect appeared much weaker than that of the value of $\rho$ (or that of the selected fitting range) on the Avrami exponent.
Specifically, plotting the time evolution of individual terms of Eqs~\eqref{FreeEnergyEq} and \eqref{eq:adim_allen_cahn} integrated over the entire domain shows that the $\widetilde{\Delta f}$ terms are dominant by nearly two orders of magnitude against the Laplacian terms during most of the simulation.
Moreover, Figs~2F and 3F, which are representative of the overall behavior of all simulations for both $\rho=1.1$ and $\rho=2.2$, show that the particles remaining closely circular, even during impingement, except in very localized regions.
While that aspect may deserve further exploration for a broader range of cases and parameters, this observation seems to indicate that the curvature plays a minor role in the kinetics of the transformation.

%%%%%%%%%%%%%%%%%%%%%%%%%%%%%%%%%%%%%%%%%%%%%%%
\section{Conclusion}\label{sec:conclusion}
%%%%%%%%%%%%%%%%%%%%%%%%%%%%%%%%%%%%%%%%%%%%%%%

We used a computationally-efficient (Julia-GPU) implementation of a simple phase-field nucleation benchmark \cite{wu_phase_2021} in order to (i) produce a statistically significant data (over hundreds of simulations and thousands of particles), (ii) analyze the kinetics of transformation, and (iii) compare the results to the classical, and widely used, JMAK theory.

Simulations of individual particles (\emph{Problem~I}) exhibited an excellent agreement to classical nucleation theory in terms of critical nucleation radius ($r^*$).
They also highlighted the increasingly high incubation time -- before the establishment of a steady growth velocity -- as the initial particle radius ($r_0$) gets close to $r^*$.

Multi-particle PF simulations of nucleation and growth generally capture JMAK theory exponent when the nuclei size $r_0$ is sufficiently higher than the critical radius $r^*$, for both site-saturated (\emph{Problem~II}) and continuous nucleation (\emph{Problem~III}) configurations. 
As $r_0\to r^*$, the increase of incubation time slows down the average transformation kinetics and the JMAK assumption of constant transformation rate is not satisfied. 
The statistical distribution of transformed fraction and that of fitted Avrami exponent are broader in the case of continuous nucleation than in the case of site-saturation, which we attribute to the greater degree of stochasticity when nuclei are seeded at random times.
We noted a significant dependence of the Avrami exponent to the selected time range for the linear fit of $\log[-\log(1-Y)]$ versus $\log(t)$ and thus suggest relying on a range of transformed fraction instead of a time range.
Finally, within the scope of the simulations and parameters considered here, we did not observe any indication of a substantial effect of interface curvature -- naturally present in PF simulations but absent from JMAK theory -- upon the transformation kinetics. 

In summary, we presented an additional illustration of the excellent agreement between PF and classical nucleation and JMAK theories, as long as original assumptions -- in particular the constant growth velocity in an infinite domain -- are reproduced. 
We also proposed a statistical picture of transformed fraction and Avrami exponents in first-order phase transformation kinetics, and of the deviation from JMAK theory for low values of $\rho$, linked to the significant incubation time and breakdown of the constant growth velocity assumption. 
A possible perspective to extent this study could be to quantitatively assess the dependence of the incubation time or early transient growth kinetics as a function of $\rho$ in order to incorporate it within the fitted law -- e.g. using the more general time cone analysis \cite{cahn_time_1996}.
While the current study is arguably a first step in the statistical analysis of computational PF results for phase transformations, it is our hope that it can encourage the community to address such stochastic problems from a statistical perspective.

%%%%%%%%%%%%%%%%%%%%%%%%%%%%%%%%%%%%%%%%%%%%%%%
\section*{Acknowledgments}
%%%%%%%%%%%%%%%%%%%%%%%%%%%%%%%%%%%%%%%%%%%%%%%

JM acknowledges support from Texas A\&M University Department of Materials Science and Engineering and the Mar\'ia de Maeztu seal of excellence of IMDEA Materials Institute from the Spanish Research Agency (CEX2018-000800-M).
VA acknowledges the support from NSF under Grant No. DMR 1905325. 
RA acknowledges the support from NSF through Grant No. NSF-CDSE-2001333. 
DT acknowledges support from the Spanish Ministry of Science through a Ram\'on y Cajal Fellowship (RYC2019-028233-I).
Part of this research was conducted with the advanced computing resources provided by Texas A\&M High Performance Research Computing.

%%%%%%%%%%%%%%%%%%%%%%%%%%%%%%%%%%%%%%%%%%%%%%%
\bibliography{AvramiPaper}
%%%%%%%%%%%%%%%%%%%%%%%%%%%%%%%%%%%%%%%%%%%%%%%

\end{document}